\documentclass[11pt]{article}

\usepackage[utf8]{inputenc}
\usepackage[T1]{fontenc}
\usepackage{textgreek}
\usepackage{amsmath,amsfonts,amssymb}
\usepackage{graphicx}
\usepackage{booktabs}
\usepackage{multirow}
\usepackage{subcaption}
\usepackage{url}
\usepackage{hyperref}
\usepackage{xcolor}
\usepackage{tikz}
\usepackage{eso-pic}
\usepackage{geometry}
\usepackage{comment}
\usepackage{natbib}
\usepackage{setspace}

\AddToShipoutPictureBG{%
\begin{tikzpicture}[remember picture,overlay]
\node[rotate=45,scale=4,text=gray!20] at (current page.center) {AUTHOR MANUSCRIPT};
\end{tikzpicture}%
}

\usepackage{tabularx}
\usepackage{dcolumn} 
\newcolumntype{d}[1]{D{.}{.}{#1}}

\usepackage{todonotes}
\let\xtodo\todo
\renewcommand{\todo}[1]{\xtodo[inline,color=green!50]{#1}}

\hypersetup{
    colorlinks=true,
    linkcolor=blue,
    filecolor=magenta,      
    urlcolor=cyan,
    citecolor=blue,
}

\title{Moderating Role of Presence in EEG Responses to Visuo-haptic Prediction Error in Virtual Reality}

\author{
Lukas Gehrke, 0000-0003-3661-1973$^{1,*}$ \and 
Leonie Terfurth, 0000-0001-6143-4222$^{1}$ \and 
Klaus Gramann, 0000-0003-2673-1832$^{1}$ \\[0.5em]
\small
$^{1}$Technische Universität Berlin, Berlin, Germany \\
\small
$^{*}$Email: lukas.gehrke@tu-berlin.de
}

\date{\today}

\begin{document}

\maketitle

\begin{abstract}




Virtual reality (VR) can create compelling experiences that evoke presence, the sense of ``being there.'' However, problems in rendering can create sensorimotor disruptions that undermine presence and task performance. Presence is typically assessed with post-hoc questionnaires, but their coarse temporal resolution limits insight into how sensorimotor disruptions shape user experience. Here, we combined questionnaires with electroencephalography (EEG) to identify neural markers of presence-affecting prediction error in immersive VR. Twenty-five participants performed a grasp-and-place task under two levels of immersion (visual-only vs.~visuo-haptic). Occasional oddball-like sensorimotor disruptions introduced premature feedback to elicit prediction errors. Overall, higher immersion enhanced self-presence but not physical presence, while accuracy and speed improved over time irrespective of immersion. At the neural level, sensorimotor disruptions elicited robust event-related potential effects at FCz and Pz, accompanied by increases in frontal midline $\theta$ and posterior $\alpha$ suppression. Through source analyses localized to anterior-- and posterior cingulate cortex (ACC/PCC) we found that PCC $\alpha$ activity showed heightened sensitivity to disruptions exclusively in visuo-haptic immersion. Exploratory moderation analyses by presence scores revealed no consistent patterns. Together, these results suggest that higher immersion amplifies both the benefits and costs of sensorimotor coherence.

\end{abstract}


\section{Introduction}

Sense of presence, the VR experience of ``being there'', is notoriously difficult to measure continuously and noninvasively. The dominant approach to measuring presence relies on post-hoc self-report questionnaires~\citep{Slater1994-sh, Witmer1998-ew, Makransky2017-hr, Schubert2003-sq}, which interrupt the experience and fail to capture moment-to-moment fluctuations~\citep{Schwind2019-ar, Slater2003-gg, Weech2019-pm}. 

Theoretical frameworks suggest that presence depends on the brain's ability to maintain coherent sensorimotor contingencies (SMCs)~\citep{Slater2009-au, Seth2014-fv}. When predictions fail to match incoming sensory signals, prediction errors (PE) arise and trigger an update of the internal model, shaping both behavior and subjective experience~\citep{Seth2014-fv, Seth2021-io}. Classic illusions such as the rubber hand demonstrate how multisensory conflicts can alter body ownership~\citep{Botvinick1998-iw}, illustrating how the brain minimizes incoherence by realigning sensory streams. From this perspective, PE provide a window into how the brain constructs a stable sense of self and environment~\citep{Tsakiris2017-oe, Chancel2022-ih}. 

In VR, these mechanisms are continuously challenged by artificial mappings between movement and sensory feedback. 
The technical features of a system determine its level of \textit{immersion}, defined as the extent to which multiple sensory channels and motor contingencies are reproduced~\citep{Slater2009-au}. Immersion thus provides the technological substrate, while presence reflects the resulting psychological state of being within the virtual environment. Higher immersion, such as through added haptic feedback, strengthens the coupling between action and perception but also increases the demand for coherence, since when the environment behaves inconsistently, the cost of disruptions may scale with immersion. 

For example, if discrepancies exceed a temporal binding window for multisensory integration~\cite{Kording2007-xf}, the system fails to integrate across modalities, triggering error signals that undermine agency and ownership; key components of the sense of embodiment that have been shown to contribute to presence~\citep{Kilteni2012-gm}. From a predictive coding perspective, such temporal misalignments can be understood as PE that exceed the tolerance of the current generative model, prompting re-estimation of the causal structure of the environment~\citep{Friston2010-hy, Limanowski2022-ds}. In our experiment, such temporal mismatches were operationalized as brief visuo-haptic glitches that violated participants' predictions about object contact.

EEG offers a promising avenue for tracking such processes with high temporal precision. Prior work has shown that PE in VR interactions elicit robust event-related potentials (ERPs) and oscillatory dynamics, providing a temporally precise proxy for disruptions in SMCs and, potentially, presence~\citep{Gehrke2019-og, Gehrke2022-tj, Singh2018-qi, Singh2021-qc, Si-mohammed2020-ru}. Controlled sensorimotor ``glitches'', such as visuo-haptic mismatches, provide systematic probes for eliciting these neural signatures~\citep{Krol2020-lj}. However, to establish EEG as a reliable and generalizable tool, it is essential to identify the neural sources that generate these signals~\citep{Michel2019-bb}. 

Two regions are particularly relevant in this respect. The anterior cingulate cortex (ACC) has been consistently implicated in monitoring unexpected outcomes, conflict, and PE. Frontal midline $\theta$ oscillations, often attributed to ACC sources, are widely regarded as a neural correlate of cognitive control~\citep{Cavanagh2014-mm}. Their phase-locked ERP counterpart, the error-related negativity (ERN)~\citep{Van_Noordt2016-qz}, was initially interpreted as a negative reinforcement learning signal conveyed via the mesencephalic dopamine system~\citep{Holroyd2002-in}. More recent perspectives emphasize a broader function, with frontal $\theta$ indexing the need for control across conflict, errors, and unexpected feedback~\citep{Cavanagh2014-mm}, and computational models proposing that the ACC computes a hierarchical, valence-agnostic signal of surprise~\citep{Alexander2019-uf}. Importantly, visuo-haptic disruptions in VR reliably elicit PE negativities measured at fronto-central electrode locations (FCz), accompanied by frontal $\theta$ increases, with source analyses implicating medial frontal generators~\citep{Gehrke2019-og, Gehrke2022-tj, Gehrke2024-xq}. 

Posterior regions, in contrast, support multisensory integration and the construction of coherent body representations within peripersonal space~\citep{Makin2007-ng}. Posterior midline structures such as the posterior cingulate cortex (PCC) have also been implicated in self-referential and body-related processing, with recent evidence suggesting that PCC activity contributes to causal inferences about body ownership~\citep{Tsakiris2017-oe, Chancel2022-ih}. Posterior $\alpha$ oscillations are central to this process, since suppression of $\alpha$ power facilitates multisensory binding~\citep{Klimesch2012-ni}, and body-illusion paradigms show that stronger $\alpha$ suppression accompanies stronger ownership experiences~\citep{Chancel2022-ih}. Converging evidence further suggests a causal role of parietal $\alpha$, as suppression at the PCC predicts visuo-tactile integration~\citep{Misselhorn2019-bs}, with parietal generators implementing Bayesian causal inferences about embodiment~\citep{Rossi_Sebastiano2024-rc}, and manipulations of $\alpha$ rhythms directly altering the stability of body ownership~\citep{D-Angelo2025-ap}. 

Together, these findings suggest a hierarchical architecture of prediction-error processing in which frontal and posterior rhythms play complementary roles, with frontal $\theta$ signaling mismatches and the need for control, and posterior $\alpha$ supporting the integration of multisensory information into a coherent body representation. 
A coherent sense of presence thus depends on the brain’s ability to maintain temporal alignment across modalities, and when this binding window is exceeded, integration fails, undermining both agency and presence. 

To test these mechanisms, we combined EEG with presence questionnaires during a visuo-haptic grasp-and-place task using a haptic glove. Participants had to grasp a ball and place it into a larger sphere in either visual-only or visuo-haptic VR. In both scenarios, grasping occasionally glitched, providing sensory feedback unexpectedly. 
This design allowed us to test three hypotheses: (i) increasing immersion through additional haptic feedback enhances subjective presence, (ii) PE elicit robust ERN and oscillatory signatures (frontal $\theta$ and posterior $\alpha$) at both sensor (FCz, Pz) and source (ACC, PCC) levels, and (iii) these neural responses scale with the number of sensory displays and inter-individual differences in presence. By linking subjective reports with temporally precise neural markers, we aimed to advance toward continuous and unobtrusive indices of presence disruptions that could ultimately support adaptive VR systems~\citep{Gehrke2025-ma}.

\section{Methods}
The present study aimed to investigate how neural markers of PE relate to the sense of presence in immersive VR. Participants performed a repeated precision grasping task at a virtual table, reaching out to pick up and place objects. Two environments differing in immersion were compared: a visual-only baseline and a visuo-haptic condition that provided additional vibrotactile and force-feedback through a haptic glove. This manipulation allowed us to examine how increasing immersion, by adding haptic contingencies, modulates presence and the neural dynamics of sensorimotor PE. 

To probe transient disruptions in sensorimotor coherence, we employed an oddball-like paradigm based on a visuo-haptic congruency manipulation. In 25\% of trials, SMCs were violated by introducing brief ``glitches'', in which visual and haptic feedback were rendered prematurely relative to the expected moment of object contact. In the remaining 75\% of congruent trials, feedback occurred as expected at contact, allowing us to contrast coherent and incoherent action-effect mappings within a continuous interaction task.

\subsection{Participants}
Twenty-five participants (17 female, 8 male; $M$ = 28.7 years, $SD$ = 4.4, range = 23--39) took part in the study. They were recruited through an online recruitment tool\footnote{\url{www.tu-berlin.sona-systems.com}} and local advertisements. All participants reported being right-handed, fluent in German (C1; $n$ = 24) or English (B2; $n$ = 1), and having normal or corrected-to-normal (contact lenses) visual acuity. Regarding prior VR experience, two participants reported no previous experience, 17 reported little experience ($\leq 5$ interactions), and six reported more extensive experience ($>5$ interactions). When asked specifically about prior experience with haptic interfaces in VR, 14 reported no experience, 10 reported little experience ($\leq 5$ interactions), and one reported more extensive experience ($>5$ interactions). Prior to the study, participants were fully informed about the nature of the experiment, the data collection procedures, and the anonymization process, and they provided written informed consent. Participants received either monetary compensation (€12/hour) or course credits. The experiment was approved by the Ethics Committee of the Department of Psychology and Ergonomics at TU Berlin\footnote{Tracking number: BPN\_GEH\_220421}.

For one participant, the experiment was terminated early due to significant challenges with depth perception, which made it difficult for them to complete the task. Consequently, the experiment was stopped after one block, and the participant was excluded from all further analyses.

\subsection{Apparatus}
\subsubsection{VR and Eye Tracking}
The VR was designed in Unity3D (Unity Software Inc., version 2019.1.8) and presented through the HTC Vive Pro Eye (High Tech Computer Co.)\footnote{VIVE Pro Eye Specs: www.vive.com/us/product/vive-pro-
eye/specs/}. Eye-tracking was collected using the built-in eye tracker (Tobii AB, version Core SW 2.16.4.67) with a sampling frequency of 120 Hz. To mask potentially distracting sounds from the environment, participants were equipped with in-ear headphones playing white noise.


\begin{figure}[!h]
  \centering
  \includegraphics[width=\columnwidth]{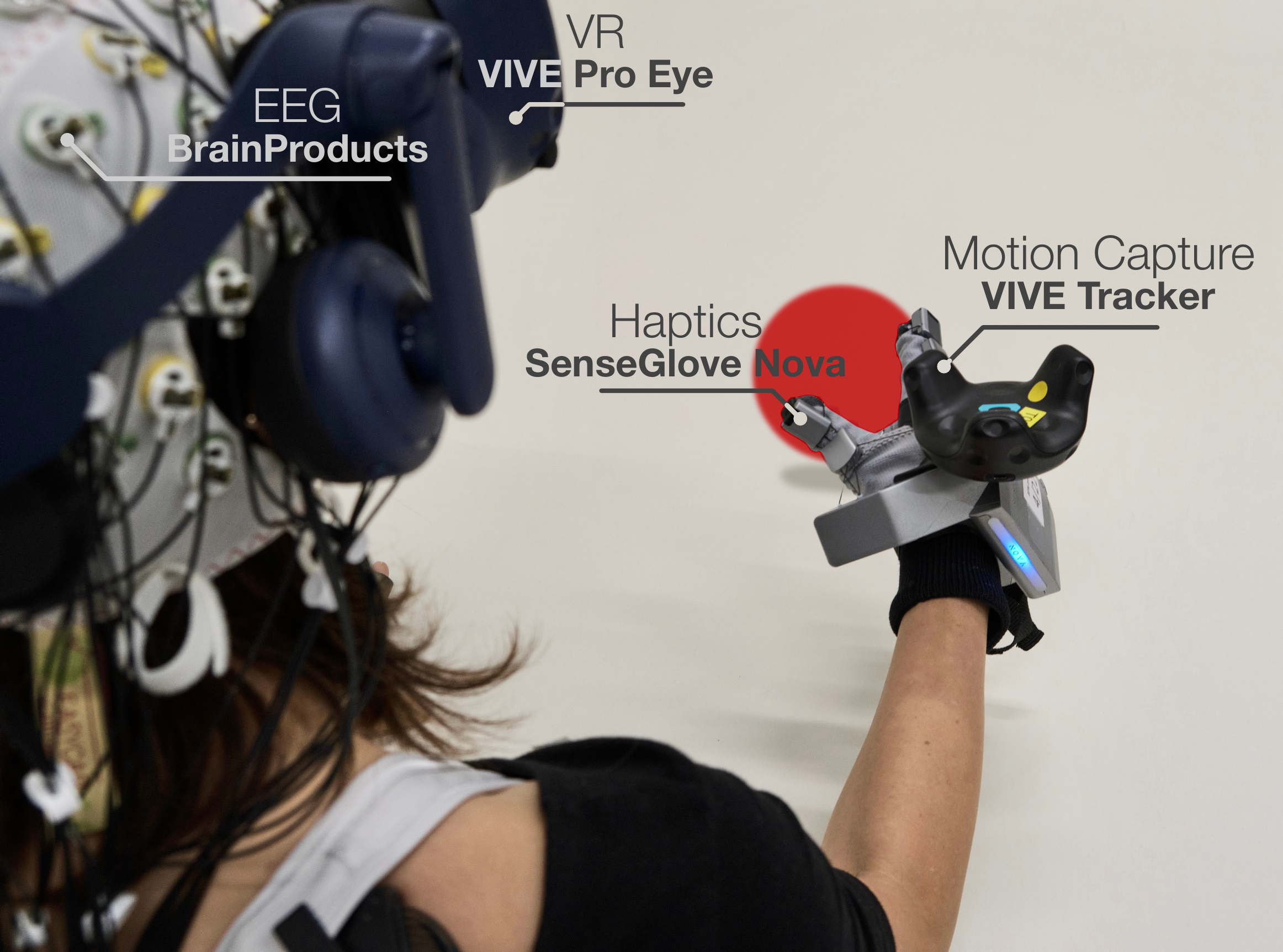} 
  \caption{Experimental setup. EEG was recorded using a 64-channel BrainProducts amplifier system. Haptic feedback was delivered through the SenseGlove Nova interface. The virtual environment was presented via the HTC VIVE Pro Eye headset. Hand position and movement were tracked using a VIVE Tracker.}
  \label{fig:setup}
\end{figure}

\subsubsection{Motion Tracking}
Finger tracking and presentation of haptic feedback were realized with the SenseGlove Nova (SenseGlove BV)\footnote{SenseGlove Nova Specs: https://senseglove.gitlab.io/SenseGloveDocs/nova-glove.html}. We used a VIVE tracker to capture the position of the hand. For the purpose of this study, we slightly adapted the functionality of the glove: First, the haptic feedback was reduced to the thumb and the index finger to limit the chance of undesirable feedback. Second, the colliders used for force-feedback were optimized for the grasping functionality (their size was adapted through trial-and-error).


\subsection{Task \& Recordings}
\begin{figure*}[!t]
  \includegraphics[width=\textwidth]{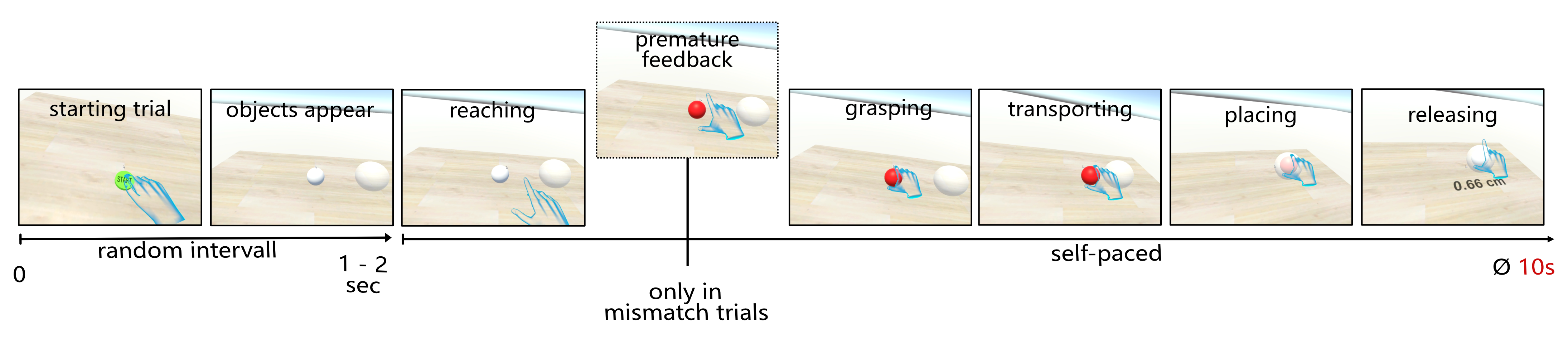}
  \caption{Task sequence. Each trial began by pressing a start button, after which one or more objects appeared following a random interval (1–2~s). Participants reached toward the object, grasped it with a precision grip, transported it to the target location, placed it, and released it. In mismatch trials, premature visual and/or haptic feedback was presented during the reaching phase. The task was self-paced and lasted on average 10~s per trial.}
  \label{fig:task}
\end{figure*}

In our study, participants repeatedly performed a precision grasping task: They placed their right hand at a resting position which triggered a green button to appear on the table. Tapping the green button then started the trial. Upon pressing the start button, and following a random variable delay of 1.5 -- 2.5 s, a white ball appeared on the table as the target object to pick up, accompanied by a target placement location (white transparent sphere\footnote{The transparent nature of the placement location was chosen to avoid feedback expectations during placement.}), see figure \ref{fig:task}. Where the target object appeared was counterbalanced between three locations on a center-line in front of the participants. The furthest distance was calibrated to the reach of the participant's outstretched arm (middle distance = - 3 cm, close distance = - 6 cm). The target placement location appeared permuted to either the left or right of the object at a distance of 30 cm.

Participants were instructed to reach for and then grasp the object with a pincer grasp (thumb and index finger extension). Participants received sensory feedback when the object was successfully grasped, see figure~\ref{fig:task} for an illustration. Next, participants were tasked to place the object at the placement location. Participants were instructed to be as accurate as possible while keeping a good pace. To keep up motivation, the accuracy, i.e., the Euclidean distance between the center of the object and the placement sphere in cm, was displayed on the table after placement. Finally, in preparation for the following trial, the object and the placement sphere disappeared 2 s after placing the object. This was also the case when the object was dropped while carrying it. To move on to the next trial, participants were required to reposition their hand back to the designated resting spot before tapping the green start button to start the next trial.

\subsubsection{Feedback Manipulations}
We manipulated the sensory feedback upon grasping the object in two ways: regarding (1) feedback congruency, i.e. oddball-like behavior, and (2) haptic immersion.

\paragraph{Feedback Congruency.} 
To induce transient sensorimotor PE, 25\% of trials (\textbf{Mismatch}) introduced brief visuo-haptic glitches. In these trials, feedback was rendered slightly before the hand made contact with the virtual object, violating participants' learned SMCs. The remaining 75\% (\textbf{Match}) of trials maintained congruent feedback timing, providing a stable baseline of coherent interaction. The feedback was timed by a velocity-based algorithm. At the peak velocity of the participant's reaching movement, the feedback was triggered and lasted for 200ms. The match to mismatch trial ratio was chosen in alignment with previous work~\cite{Gehrke2022-tj, Gehrke2019-hp}. To avoid potential bias due to sequence effects of feedback congruency, mismatch and match trials were fully randomized.

\paragraph{Immersion.}
Two immersion levels were compared: In the \textbf{visual-only condition}, a successful grasp interaction was indicated by a color change (white to red) of the grasped object. In visual-only mismatch trials, the premature feedback was displayed as a short red flicker of the object, see figure~\ref{fig:task}. In the \textbf{visuo-haptic condition}, successful grasping of the ball was indicated by visual-, vibro-tactile and force-feedback. Here, vibro-tactile feedback simulated the touch of the objects' surface as short vibrations on the tip of the thumb and the index finger. Force-feedback rendered the physical resistance when holding an object between the fingers by blocking of thumb and index finger extension via the glove interface's brake mechanism. In visuo-haptic mismatch trials, all feedback modalities were presented at the mismatch time. 

In both levels of immersion, the grasping was realized by the finger tracking capabilities of the glove interface. 

\subsubsection{Procedure}
The experimental procedure started with a setup and calibration phase which included the fitting of the psychophysiological measuring instruments, the haptic glove, and the VR headset. We calibrated the haptic glove to each individual participant with the software provided by the manufacturer. Next, the height of the virtual table and the distance of the objects were calibrated to the participants sitting position and reach distance. Prior to fitting the VR glasses and calibrating the eye tracker (5-point calibration), participants were introduced to the grasping task by performing an exemplary task sequence outside of VR using a tennis ball. After putting on the VR glasses, participants were presented further task instructions inside the VR scene. The task was framed by telling participants that interactions in VR are still in a developmental stage and therefore, they might encounter some inaccuracies. They were told that when that happens they should just continue with the task.

To familiarize participants with the hardware and assess whether haptic realism impacts presence experience, the experiment started with two \textit{training} blocks with 15 trials in each condition (visual-only and visuo-haptic). When participants struggled with the grasping mechanism during the training phase, they were asked to perform a few more trials. To support the grasping success, we asked them to try different grasping techniques and suggested different grasping options. At the end of each \textit{training} block, a presence questionnaire was presented in the VE~\cite{Feick2020-fb}. We used the sub-scales physical presence and self presence from the Multidimensional Presence Scale~\citep{Makransky2017-hr}.

Following the two \textit{training} blocks, participants completed four experimental blocks, each containing 76 trials. After the third and fourth experimental block, the presence questionnaire was administered again. Between blocks, participants took a break of at least 60 seconds. The immersion condition of the blocks (visual-only, visuo-haptic) alternated, with the starting immersion counterbalanced between participants. At the end of the experiment, participants were interviewed. This exit interview contained open questions exploring whether participants noticed the implemented congruency manipulation (mismatch trials) and the differences in immersion (visual vs. visuo-haptic). 


\subsubsection{Data Records: ExG and Motion Capture}
We recorded EEG from 64 actively amplified wet electrodes using BrainAmp DC (BrainProducts GmbH) amplifiers. Electrodes were placed in elastic caps according to the extended 10\% system~\citep{Oostenveld2001-yd}. All data were sampled at 250 Hz with FCz as the reference channel and Fpz as ground electrode, an analog highpass filter of 0.0159 Hz (10-second time constant), and a low pass determined by half the sampling rate. After fitting the cap, all electrodes were filled with conductive gel and electrode impedance was brought below 5k$\Omega$ where possible. One electrode below the right eye was used to record vertical eye movements. Electrodermal activity (EDA) and the elctrocardiogram (ECG) were recorded with an ExG amplifier (BrainProducts GmbH). We placed the EDA electrodes on the ring and middle finger of the left hand after applying isotonic gel. We acquired the ECG signal using lead II of Einthoven's triangle with a positive electrode on the left leg, a negative electrode on the right arm, and a reference electrode on the right leg. All physiological recordings were sampled at 250 Hz.

All continuous data streams and an experiment marker stream were recorded and synchronized using LabStreamingLayer~\cite{Kothe2025-oh}. See figure \ref{fig:setup} for the full experimental setup.

\subsubsection{EEG Signal Processing}
EEG data were preprocessed using the BeMoBIL-pipeline, wrapping and extending EEGLAB toolboxes~\cite{Delorme2004-sn, Klug2022-lc} running in a MATLAB 2023b environment (The MathWorks Inc., USA). The inherent delay of the BrainVision EEG setup using LSL was corrected by subtracting 60~ms from the timestamps\footnote{https://wiki.bpn.tu-berlin.de/wiki/doku.php?id=lab:lab\_software:lsl:lsl-test, last accessed 13/9/2022}. After removing non-experiment segments, bad channels were detected using the `FindNoisyChannel' function, which selects bad channels by amplitude, the signal to noise ratio and correlation with other channels~\cite{Bigdely-Shamlo2015-ds}. Rejected channels (mean: 8.42, SD: 10.14) were then interpolated while ignoring the EOG channel, and finally re-referenced to the average of all channels, including the original reference channel FCz. After applying a high-pass filter at 1.5~Hz, the EEG data were decomposed using independent component analysis (ICA)~\cite{Makeig1995-cf}. The AMICA algorithm was used with its automatic time-domain cleaning~\cite{Palmer2011-zs}. Subsequently, equivalent dipole models were fitted to the resultant independent components (ICs) using a boundary-element head model. ICs reflecting eye movement were removed based on the classification from the ICLabel toolbox~\cite{Pion-Tonachini2019-fy}. For this, ICLabel's popularity classifier was used, meaning that all components having the highest probability for the eye class were projected out of the sensor data.

In the first step to prepare the cleaned data for analyses, noisy (extremely large amplitude fluctuations) trials were rejected. To this end, the EEGLAB function \textit{autorej} was used on EEG epochs -1 to 2~s around the grab event, keeping the default parameters: a voltage threshold of 1000~$\mu$V, a standard deviation cutoff of 5, a maximum rejection rate of 5\% per iteration, and all channels included. This function iteratively rejects epochs with improbable data based on standard deviation thresholds, adaptively increasing the threshold when too many epochs exceed the limit. 


\paragraph{Source Localization \& Group-level Clustering} 
\label{ic_clustering}
For group-level analyses of IC activity, independent components (ICs) were retained if their ICLabel probability of being brain-related exceeded 0.20 and their probability of being channel noise was below 0.05. For clustering, we followed an ROI-driven, repetitive k-means clustering approach (cf.~\citet{Gramann2021-ug}).


To avoid circularity with subsequent ERSP analyses, preclustering relied exclusively on anatomical information, using equivalent dipole locations (weight = 1), while spectral, ERSP, scalp, and ERP features were excluded (all weights = 0). In total, 467 ICs entered the clustering, corresponding to a mean of 19.5 (SD = 5.5) per participant. The number of clusters was set adaptively to seventy percent of the average IC count per subject, yielding $k = 14$. Outliers exceeding 2.5 standard deviations from any centroid were excluded. 

We then performed repeated clustering with 2000 iterations. The final solution was chosen using a weighted quality score that emphasized broad participant representation, compact clusters, low dipole residual variance, and proximity to the ROI. To test anatomical hypotheses, clustering was run twice to optimize for two a priori regions of interest: the anterior cingulate cortex (ACC; MNI coordinates $x=0$, $y=30$, $z=20$) and the posterior cingulate cortex (PCC; MNI coordinates $x=0$, $y=-55$, $z=45$). 

The final solution for the ACC cluster comprised 28 ICs from 21 participants (ratio = 1.33). The mean residual variance (RV) was $10.9\%$, with a spatial spread of $20.3$~mm and a centroid distance of $20.2$~mm from the ROI coordinates in Talairach space. The PCC cluster comprised 24 ICs from 19 participants (ratio = 1.26). The mean RV was $6\%$, with a spread of $16.5$~mm and a distance of $4$~mm from the ROI coordinates. 

If multiple ICs from the same participant contributed to a given cluster, we computed a participant-wise weighted average of the power values for the subsequent analyses. ICLabel brain-class probabilities were used as weights, normalized such that $\sum_i w_i = 1$ for each participant.

\subsection{Hypotheses Testing}
First, we hypothesized that immersion would influence the sense of presence and participants' performance. Second, we examined whether PE arising from the congruency manipulation would elicit distinct event-related potentials (ERPs) and whether these neural responses would be moderated by the sense of presence of participants. Third, we hypothesized that task-related oscillatory dynamics, particularly in frontal theta and parietal alpha bands, would vary with both interaction difficulty and subjective presence, reflecting cognitive control and attentional engagement.

To address these aims, several analyses were conducted. Subjective presence scores were analyzed using inferential statistics to quantify differences between feedback conditions. EEG feature extraction focused on time-locked responses to PE (ERP at mismatch onset) and holistic motor and corrective processes (ERSP for the full grasp epoch). These neural features were then analyzed in relation to the experimental manipulations and regressed against individual presence scores to uncover potential modulation by subjective experience.

To ensure data quality, outliers were removed prior to statistical analyses. Trials were excluded if (i) they were rejected based on EEG quality (see above description of \textit{autorej} function), (ii) total trial duration exceeded 10~s, (iii) grasp duration was longer than the participant-specific median plus 0.5~s, or (iv) the mismatch occurred implausibly early (earlier than the median minus three median absolute deviations). On average, 37.60 trials per participant were excluded due to excessively long or noisy epochs (SD = 8.91), 56.52 trials per participant were excluded due to unusually long grasp durations (SD = 23.22), and 2.76 trials per participant were excluded as implausibly early mismatches (SD = 2.37). In total, these criteria led to the removal of approximately 96.88 (out of a total of 334) trials per participant. These deviations were expected given the novelty of the force-feedback functionality and underline the relevance of this research. While such trials are inherently interesting and may provide valuable insights into real-world interaction errors, they were excluded here to maintain a clear focus on the originally conceived experimental manipulations. To finalize the design matrix, all continuous predictors were z-scored within participants to stabilize variance.

\subsubsection{Sense of Presence \& Behavior}
Three measures were assessed to capture different aspects of participants' subjective experience and performance. Besides the presence questionnaire scores, `accuracy', reflected task performance quality, and `trial duration' served as an indicator of interaction fluency and efficiency.

\paragraph{Presence Scores}
Questionnaire scores from both, the physical presence, and the self-presence sub-scales were analyzed. Scores were available for $n = 22$ out of the 24 participants retained in the final data set. Hence, two participants were excluded from questionnaire-based analyses due to incomplete data entries or technical issues during questionnaire presentation. Therefore, all correlation and moderation analyzes involving the below-mentioned presence scores were performed on this data subset, while all other behavioral and EEG analyses included the full sample of participants $n = 24$.

To examine how presence ratings were affected by both immersion and time, we fit a linear mixed-effects model for both sub-scale (physical presence, self presence). Here, time was a categorical predictor with two levels (training \& experiment phase), providing insight into performance changes with time on task. Hence, the fixed effects included immersion (visual-only vs.\ visuo-haptic), experiment phase (training vs.\ experimental blocks), and their interaction. A random intercept for each participant (id) was added to account for repeated measurements. The model was specified as: \[\text{presence} \sim \text{immersion} \times \text{phase} + (1|\text{id})\]

Models were fit using the \textit{Lmer} class from the \textit{pymer4} package~\cite{Jolly2018-it}, and all parameters were estimated using maximum likelihood estimation~\cite{Pinheiro2006-bk}. The significance of fixed effects was assessed via likelihood-ratio tests comparing the full model against appropriate reduced models (e.g., without interaction terms).


\paragraph{Speed and Accuracy}
Placement accuracy was defined as the Euclidean distance (in cm) between the center of the object and the center of the placement sphere at the time of object release. Trial duration was defined as the time from object spawn to successful object placement. Similar to the presence scores model, linear mixed-effects models with fixed effects for \textit{immersion} (visual-only vs.\ visuo-haptic), \textit{experiment phase} (training, experimental block 1 \& 2), and their interaction were fit for both variables. A random intercept for participant accounted for repeated measures. Each model was fit and assessed in the same way as the model for the presence scores above. 

\subsubsection{EEG}
Two EEG analyses approaches were employed, each aligned with the cognitive demands of specific trial phases. Event-related potentials (ERPs) were analyzed at the sensor level to capture transient, phase-locked responses to the occurrence of visuo-haptic mismatches compared to normal (match) trials at the moment of object grasp. This analysis served both as a validation of the experimental manipulation and as a probe of PE processing. Time–frequency dynamics (ERSPs) were analyzed for channel FCz and at the source level (IC activation) ICA is more effective in isolating sources that exhibit sustained or oscillatory activity unfolding across longer timescales. In particular, the grasp moment, especially following longer or disrupted reach trajectories, was expected to engage effortful correction and motor closure, processes that are not strictly phase-locked and therefore less evident in ERP averages, but well captured by time–frequency analysis of IC sources (e.g., frontal theta, posterior alpha).

\paragraph{ERP}
For ERP analysis, data were band-pass filtered between 0.1~Hz and 15~Hz. ERPs were time-locked to the onset of the visuo-haptic feedback, specifically to the occurrence of the designed mismatch (mismatch) at the peak velocity of the grasping movement, and compared against match (successful grasp) trials. 


For each participant and time point, an ordinary least squares (OLS) regression model was fitted with predictors \textit{congruency} (match vs.~mismatch), \textit{immersion} (visual-only vs.~haptic), their interaction, and a baseline covariate (baseline window -100 to 0~ms preceding the grasp event). The baseline covariate was z-scored to standardize scaling relative to the categorical predictors. The full model was hence specified as: \[\text{erp} \sim \text{congruency} \times \text{immersion} + \text{baseline} + 1\] This yielded subject-specific beta estimates for each effect across time. The resulting time-resolved coefficients were submitted to nonparametric one-sample $t$-tests against zero, implemented as cluster-based permutation tests with 10,000 permutations in \textit{MNE-Python}~\cite{Gramfort2013-fa}. Statistical inference was based on threshold-free cluster enhancement (TFCE) with a stepwise thresholding approach (start~=~0, step~=~0.05). Significant clusters ($p < .05$) indicate time windows in which the given predictor reliably explained variance in the ERP across participants.

\paragraph{Topography}
To visualize the spatial distribution of ERP effects, we computed scalp topographies of first-level OLS regression coefficients. For each effect of interest (\textit{congruency}, \textit{immersion}, \textit{interaction}), we extracted beta estimates at multiple latencies (100, 150, 200, 250, and 300~ms) across all participants. For each channel, a $t$-test against zero was performed, and the resulting $p$-values across channels were corrected for multiple comparisons using false discovery rate (FDR)~\cite{Benjamini1995-cw}. Channels surviving an $\alpha = .01$ threshold after correction were highlighted in the scalp maps.

\paragraph{Post-hoc exploration: Correlation with Presence}

To assess whether individual differences in subjective presence moderated EEG effects, we regressed the \emph{delta presence score} between the visuo-haptic and visual baseline \textit{experiment} blocks on participants' first-level estimates. Specifically, we tested two moderation models: (i) regressing the presence delta on the \(\beta_{\text{immersion}}\), which addresses whether participants who experienced stronger increases in presence from added haptics also showed larger neural immersion effects; and (ii) regressing the presence delta on the \(\beta_{\text{interaction}}\), which addresses whether increases in presence predicted the extent to which immersion modulated PE processing. We did not include the \(\beta_{\text{congruency}}\) in these moderation analyses, as this estimate reflects the main effect of mismatch across both immersion conditions and therefore has no conceptual link to inter-individual differences in presence that were defined by the contrast between immersion conditions. Moderation coefficients were visualized as ERP-like plots and as scalp topographies at multiple latencies (100--300~ms in 50~ms steps), with significant time-points/channels masked at an uncorrected threshold of $p<.05$.

\paragraph{Source-space Oscillatory Power}
Time-frequency representations for both the ACC and PCC cluster and all single trials were computed using EEGLAB's \textit{newtimef()} function (3-80~Hz in logarithmic frequency scaling), applying a wavelet transformation with 3 cycles at the lowest frequency and a linear increase of 0.5 cycles with each frequency bin. To account for variable trial durations, epochs were temporally realigned by piecewise-linear time warping based on task anchor events (spawn, approximate grasping movement velocity peak, grasp, placement). The warping preserved event order while compressing or expanding intervening intervals, enabling comparison of oscillatory activity across trials of different lengths.


Spectral power was first averaged within canonical frequency bands of interest (theta: 4--7~Hz, alpha: 8--12~Hz, beta: 13--30~Hz) for each trial. These band-limited raw power values were then baseline-corrected by divisive normalization relative to the pre-stimulus window ($-300$ to $-100$~ms), and the resulting ratios were converted into decibels ($10\log_{10}$). From these baseline-corrected and log-transformed values, condition-specific averages were computed, and contrasts were derived for the main effects of \textit{immersion} (haptic vs.~visual), \textit{congruency} (mismatch vs.~match), and their interaction \big((\text{haptic mismatch} - \text{haptic match}) - (\text{visual mismatch} - \text{visual match})\big). These time-resolved contrast estimates were subjected to nonparametric statistical testing using cluster-based permutation $t$-tests with $10{,}000$ permutations. As above statistical inference was based on threshold-free cluster enhancement (TFCE) with a stepwise thresholding approach (start~=~0, step~=~0.05). Significant clusters ($p < .05$) indicate time windows in which the given predictor reliably explained variance in the band power across participants. The \emph{delta presence} moderation analyses that was conducted for ERP was also applied to the band-power contrasts.




\section{Results}
We first report behavioral performance, followed by ERP and ERSP results.

\subsection{Presence \& Performance}
\begin{figure*}[!ht]
  \includegraphics[width=\textwidth]{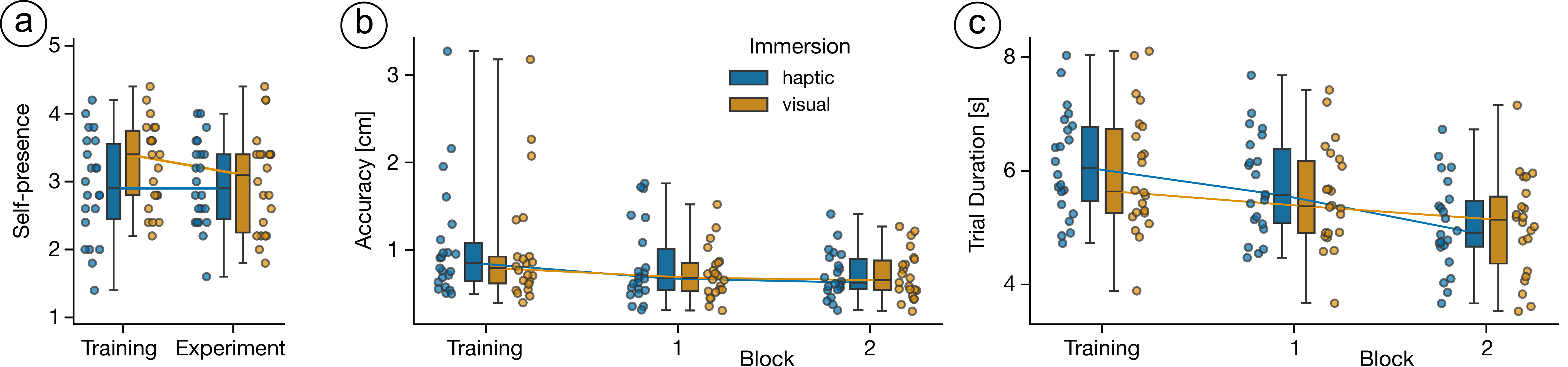}
  \caption{Behavioral results. \textbf{a:} Self-presence ratings, \textbf{b:} mean placement error (cm), and \textbf{c:} trial duration by immersion level and experiment phase. Error bars indicate the standard error of the mean.}
  \label{fig:results_behavior}
\end{figure*}

For self-presence scores, we observed a main effect of immersion, $\chi^{2}(2) = 8.34$, $p = .02$, with higher scores in the visuo-haptic compared to the visual-only condition ($\beta = 0.35$, $SE = 0.12$, $t = 2.97$; see figure~\ref{fig:results_behavior}a). There was no effect of experiment phase ($\chi^{2}(2) = 4.81$, $p = .09$) and no interaction ($\chi^{2}(1) = 3.43$, $p = .06$). No effects were observed for physical presence.

For task performance, accuracy did not differ between levels of immersion but increased over experiment phases ($\chi^{2}(2) = 24.52$, $p < .001$), as evidenced by a decreasing placement distance ($\beta = -0.21$, $SE = 0.06$, $t = -3.64$; see figure~\ref{fig:results_behavior}b). No interaction effect was observed. Similarly, trial duration did not differ between immersion but decreased over experiment phases ($\chi^{2}(2) = 58.37$, $p < .001$; $\beta = -0.41$, $SE = 0.06$, $t = -6.68$; see figure~\ref{fig:results_behavior}c).

\subsection{ERPs and Topographies}
\begin{figure}
  \centering
  \includegraphics[width=\columnwidth]{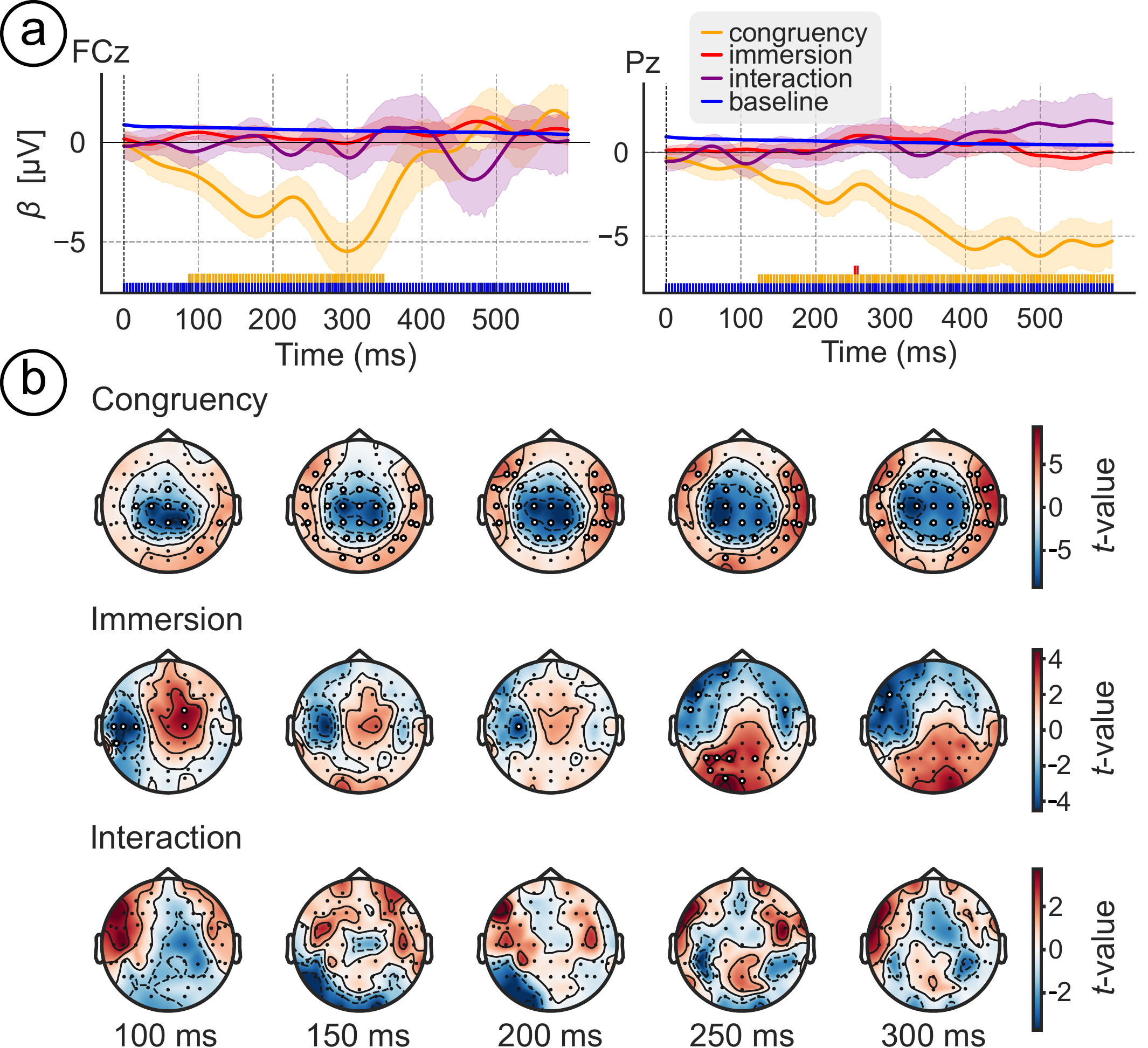}
  \caption{ERP results. \textbf{a:} Estimated ERP effects at FCz and Pz showing model-derived beta coefficients for the baseline and experimental factors (congruency, immersion, and their interaction). \textbf{b:} Scalp topographies of the experimental effects at representative latencies (100–300~ms). Enlarged electrodes indicate clusters with significant effects.}
  \label{fig:results_erp}
\end{figure}

TFCE-corrected tests of beta coefficients revealed a negative congruency effect at FCz between 100--350~ms after the grab/glitch event. The mean $\beta$ across participants at 300~ms was $-5.48~\mu\text{V}$ (95\% CI [--6.94, --4.02]), reflecting reduced amplitudes for mismatch compared to match trials (see figure~\ref{fig:results_erp}a, left). At both FCz and Pz, baseline amplitude was a significant predictor for each post-event time point. At Pz, a negative congruency effect was observed from 120~ms after the grab/glitch event until the end of the tested window. In a brief window around 250~ms post event, a positive immersion effect was observed. The mean $\beta$ across participants was $1.02~\mu\text{V}$ (95\% CI [0.41, 1.64]), reflecting increased amplitudes for haptic compared to visual-only trials (see figure~\ref{fig:results_erp}a, right).

Scalp maps revealed that for several central locations, mismatch trials had lower amplitudes than match trials (congruency effect; see figure~\ref{fig:results_erp}b, top). In contrast, electrodes on the lower left and right sides of the montage showed higher amplitudes in mismatch compared to match trials. This pattern was evident throughout 100--300~ms. For the immersion effect, a dipolar pattern was observed, with two center-right locations showing positive effects and a few central-left locations showing negative effects at 100~ms post event. At 250--300~ms, a frontal negative vs.\ posterior positive pattern emerged (see figure~\ref{fig:results_erp}b, center). No clear pattern was observed for the interaction effect (see figure~\ref{fig:results_erp}b, bottom).

\begin{figure}
  \centering
  \includegraphics[width=\columnwidth]{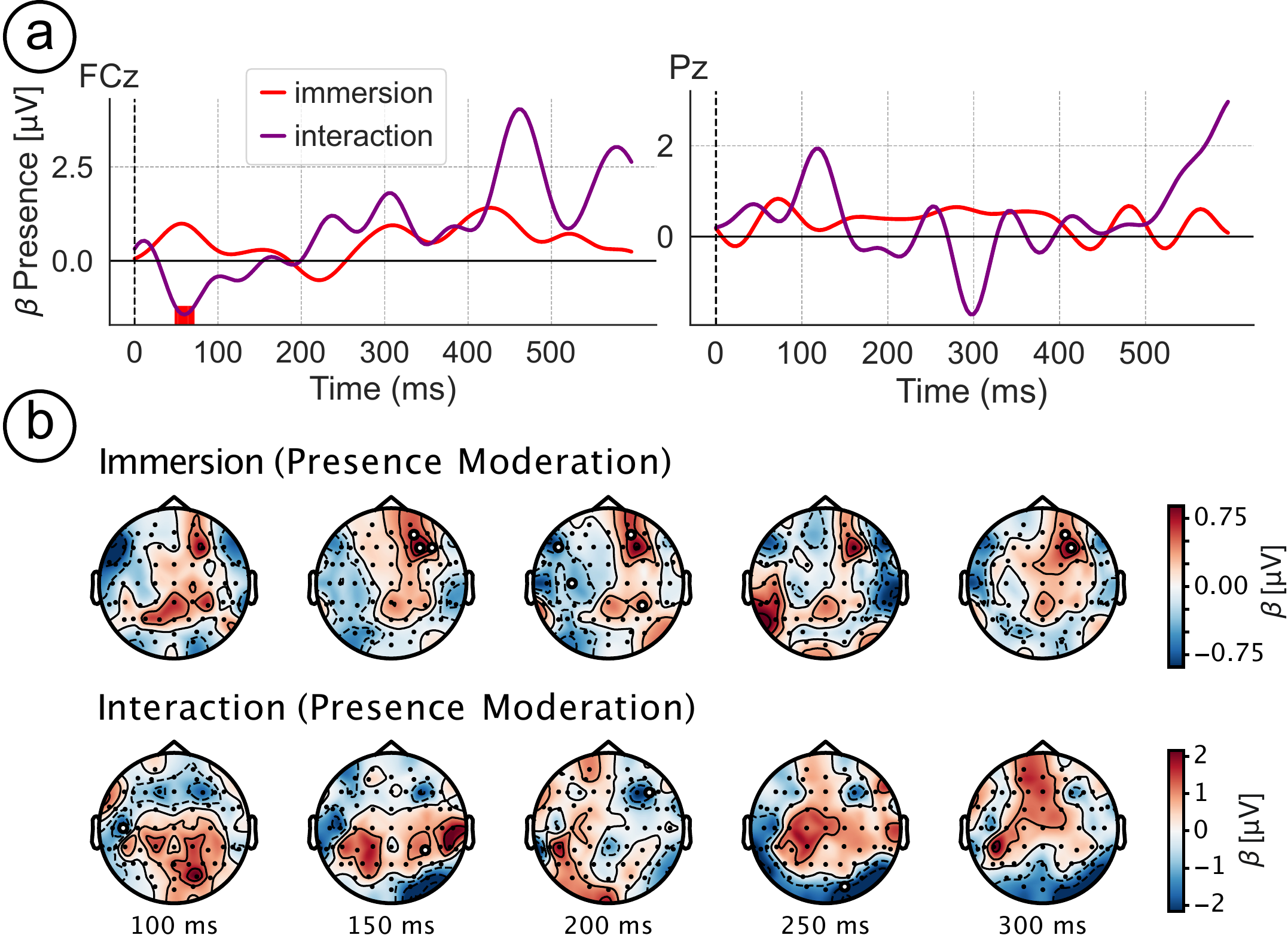}
  \caption{Moderation by presence scores. \textbf{a:} Moderation of model-derived beta coefficients at FCz and Pz by $\Delta$ presence scores. \textbf{b:} Scalp topographies of moderation effects at representative latencies (100–300~ms).}
  \label{fig:results_erp_presence}
\end{figure}

Regressing $\Delta$ presence scores between the visuo-haptic and visual-only conditions did not reveal any major effects on ERP amplitudes at FCz, Pz, or on the scalp distribution. Exploring the data at uncorrected $p$-values suggested an early moderation effect on immersion, reflecting higher amplitudes at FCz at 60--80~ms in the haptic condition for participants who reported increased presence (see figure~\ref{fig:results_erp_presence}a). This moderation effect was also apparent at a few right-frontal channels at multiple time points (150, 200, and 300~ms; see figure~\ref{fig:results_erp_presence}b, top).

\subsection{ERSP and Band Power at FCz, ACC, and PCC}
\begin{figure}
  \centering
  \includegraphics[width=\columnwidth]{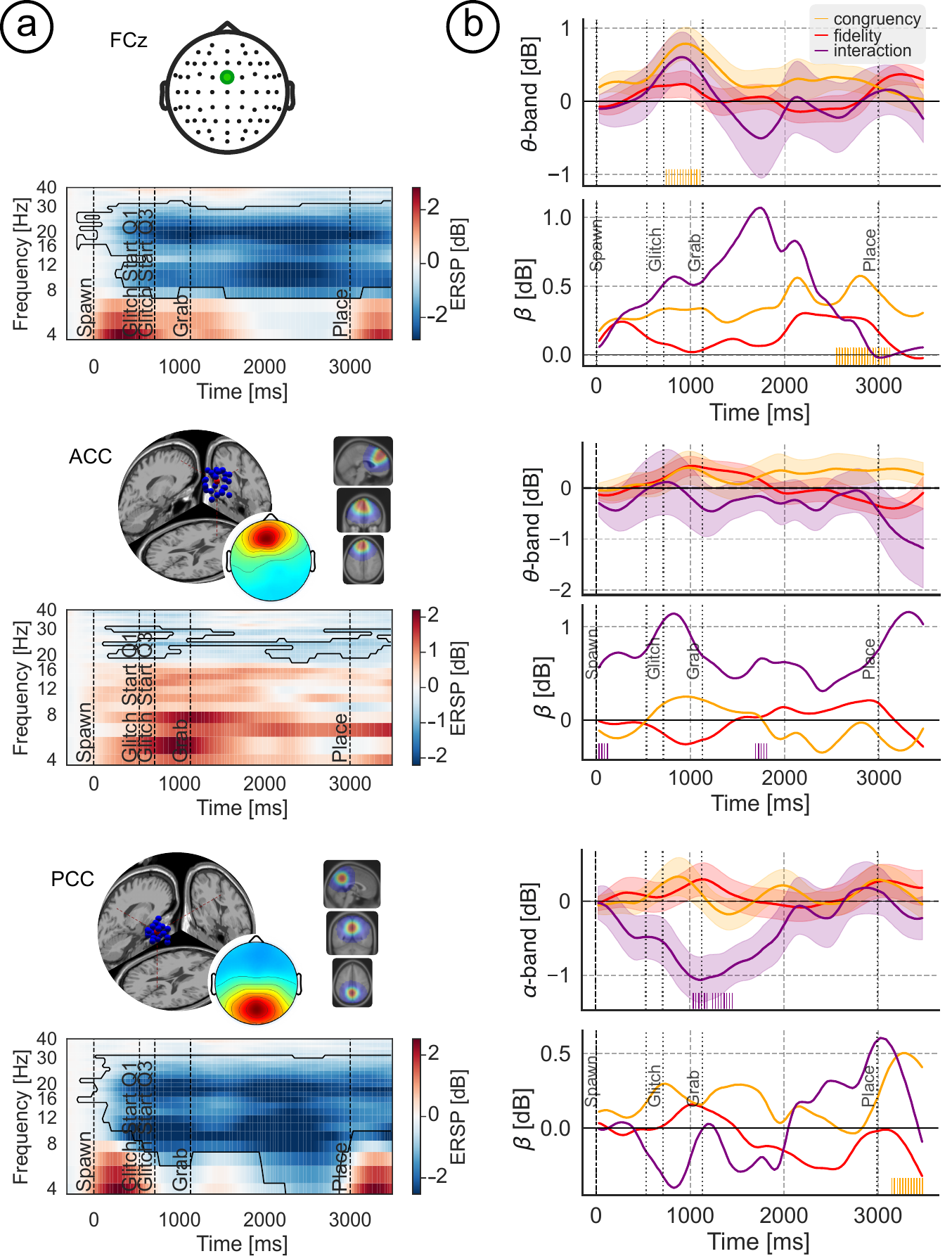}
  \caption{Time-frequency results. \textbf{a} Baseline-corrected grand-average ERSP at electrode FCz, cluster ACC and PCC, with task events aligned by temporal warping (Spawn, Glitch, Grab, Place). Black contours indicate time-frequency clusters exceeding significance from baseline according to TFCE ($p < .05$). \textbf{b} Top: Band-limited power time courses at FCz for $\alpha$- and $\theta$-bands, separated by experimental factors (immersion, congruency, interaction). \textbf{b} Bottom: Moderation of $\Delta$ presence scores on the band-power by experimental factor.}
  \label{fig:results_ersp}
\end{figure}

At electrode FCz, TFCE-corrected permutation tests ($p < .05$) revealed a sustained decrease in spectral power from 8–30~Hz between object spawn and extending past the placement event (maximum mean effect within the significant cluster = --3.42~dB) in the baseline-corrected grand-average ERSP. In addition, non-significant increases were observed in power below 6~Hz following the spawn and placement events (see figure~\ref{fig:results_ersp}a, top). In the $\theta$ band, the congruency manipulation significantly affected EEG activity at FCz in a time window following the glitch event, just preceding the grab (see figure~\ref{fig:results_ersp}b, top). Here, the maximum impact of a mismatch on $\theta$-band power was a $0.79~\text{dB}$ (95\% CI [0.35, 1.23]) change from baseline at 950~ms following the spawn event. Regressing the $\Delta$ presence scores as a moderator revealed no significant moderation effect on $\theta$-band activity at FCz (see figure~\ref{fig:results_ersp}b, top).

Grand-average IC activity in the ACC cluster exhibited a significant power decrease between 20–30~Hz following the spawn event and sustaining throughout the entire trial duration (maximum mean effect within the significant cluster = --0.99~dB). Non-significant increases in $\theta$-band power were observed following the glitch event and into the grab event (see figure~\ref{fig:results_ersp}a, center). In the $\theta$ band, the experimental manipulations did not affect EEG activity (see figure~\ref{fig:results_ersp}b, center). Regressing the $\Delta$ presence scores as a moderator indicated minor and distributed moderation effects on the interaction term in $\theta$-band activity (see figure~\ref{fig:results_ersp}b, center).

IC activity in the PCC cluster showed a sustained decrease in spectral power from 8–30~Hz between object spawn and extending past the placement event (maximum mean effect within the significant cluster = --4.1~dB) in the baseline-corrected grand-average ERSP. Non-significant brief increases in power were observed below 6~Hz following the spawn and placement events. A significant interaction effect was observed on $\alpha$-band power around the grab event, such that only in the haptic condition mismatch trials resulted in a --1.07~dB (95\% CI [--1.68, --0.46]) $\alpha$-power decrease from baseline (see figure~\ref{fig:results_ersp}b, bottom). $\Delta$ presence scores moderated the congruency effect on $\alpha$-band activity at PCC at several time points following the placement event (see figure~\ref{fig:results_ersp}b, bottom).

\section{Discussion}

In this study, we investigated how disruptions of SMCs in immersive VR environments influence performance, subjective presence, and EEG responses. Task duration and placement accuracy revealed no significant differences between immersion conditions, consistent with previous VR studies showing that behavioral metrics often remain stable despite substantial differences in immersion~\cite{Bowman2012-ga}. Higher haptic immersion did, however modestly, increase self-presence, though the overall effect might have been constrained by the impact of frequent visuo-haptic glitches. On the neural level, we replicated canonical mismatch negativities and frontal theta bursts at FCz, supporting the robustness of mobile EEG to capture error-related activity in complex VR settings. Beyond these general effects, posterior alpha suppression in the PCC emerged as selectively sensitive to visuo-haptic incongruencies. Together, these results delineate complementary roles of frontal and posterior networks: rapid frontal monitoring of prediction violations, largely agnostic to immersion, and posterior modulation of multisensory coherence, where higher immersion both enhances embodiment and increases the cost of incoherence.

\subsection{Modest Presence Gains with Increasing Immersion}

A first point of discussion concerns the modest effects of haptic immersion on presence. One potential explanation is that while additional sensory cues can strengthen embodiment, a component known to contribute to overall presence~\cite{Kilteni2012-gm}, they may also raise vulnerability to incoherence~\cite{Gonzalez-Franco2017-ff}. Indeed, our results suggest that the benefits of additional haptic immersion are counterbalanced by the heightened salience of glitches. In our case, participants may have reweighted visual over haptic input when encountering inconsistencies, leading to only incremental gains in self-presence. Furthermore, while the force-feedback capabilities of the haptic glove are a technological feat, the technology is still lacking the precision to consistently deliver a well-timed, expected haptic feedback. In such scenarios, the brain may compensate by down-weighting the haptic input and relying more on the visual input to maintain coherence. Similar observations have been made in studies on cue-weighting of haptic feedback, where the unprecise sensory channel, either haptic or visual, was down-weighted~\cite{Gibo2017-gi}. Cue-weighting, or precision-weighting in predictive processing, refers to the brain's adaptive process of adjusting the relative importance of different sensory inputs based on their reliability or noise, where less reliable channels receive reduced influence in multisensory integration, for example for haptic feedback in VR~\cite{Gall2018-pb}.

While we found a small effect of immersion on self-presence, the absence of an effect on physical presence was unexpected. Potentially, this pattern can be explained by considering the task-specific nature of our paradigm and the technological challenges inherent to the haptic glove. As evidenced by the substantial number of trials requiring removal due to prolonged grasp durations (see Appendix), participants experienced difficulties grabbing the target object in the haptic immersion condition where force-feedback was active. While we removed trials with implausibly long grasp durations from the analyses, the psychological impact may have persisted throughout the experiment. These technological limitations likely undermined participants' ability to develop stable expectations about the virtual environment's causal relationships (coherence), particularly affecting the physical presence aspects of `sense of control in the virtual environment' and `not being aware of the physical mediation'~\cite{Makransky2017-hr}.

Such challenges highlight a broader methodological issue using break-in-presence (BiP) paradigms: questionnaire-based presence ratings can become unreliable when participants experience frequent BiPs~\cite{Schwind2019-ar}. In~\cite{Gehrke2019-og}, we observed that manipulating haptic realism produced EEG PE, but little change in presence questionnaire scores, underscoring a potential limited sensitivity of subjective measures to subtle variations in haptics. One promising direction for future research lies in exploring alternative experimental paradigms. One such approach are auditory oddball paradigms that present stimuli in the real environment while participants remain in VR.~\citep{Savalle2024-co} showed that real-world auditory oddballs elicit P300 amplitudes inversely related to presence, suggesting that reduced attention to external stimuli reflects deeper immersion in the virtual world~\cite{Savalle2024-co}. These paradigms re-conceptualize BiP as attentional transitions between the real and virtual environment that do not directly affect the task at hand. Another promising approach are multimodal matching paradigms, building on Slater's sensorimotor contingency framework~\cite{Slater2009-au}. Here, participants actively explore graded sensorimotor configurations, such as varying degrees of haptic feedback, and select the configuration in which they `feel most present'. Such paradigms conceptually extend into neuroadaptive VR~\cite{Gehrke2025-ma}, where the system adaptively adjusts the feedback based on users' neural and behavioral responses to optimize coherence and presence. 

Lastly, concerning the inclusion of an initial embodiment phase: Research shows that users can embody avatars remarkably quickly, often within less than a minute once visual and proprioceptive feedback are aligned~\cite{Gonzalez-Franco2018-jz}. Establishing embodiment before the experimental manipulation could provide a stable baseline of body ownership and agency, thereby enhancing the sensitivity of presence measures to subsequent coherence disruptions. Such embodiment phases are rarely implemented in BiP paradigms but may amplify EEG sensitivity and improve ecological validity, as they ensure that predictive mechanisms operate on a fully established virtual self.

\subsection{EEG Evidence: Frontal Detection and Posterior Integration of PE}

Turning to the neural findings, our ERP analyses revealed robust PE responses across both frontal and parietal midline electrodes. At FCz, we observed two prominent negative effects of congruency peaking around 170~ms and 300~ms following the glitch/grab event, while at Pz, a sustained negative effect of congruency emerged from approximately 120~ms onward without distinct peaks. Additionally, an immersion effect emerged at Pz around 250~ms, where haptic trials showed increased positivity relative to visual-only trials. Importantly, these estimates represent model-based $\beta$ coefficients that correct for interaction terms instead of averaging across conditions which potentially inflates main effects.

The early- to mid-latency frontal negativity replicates a well-established neural signature of PE monitoring during naturalistic interaction i VR, the PE Negativity (PEN), localized to the medial frontal cortex and typically maximal at FCz $\approx$~170--250~ms~\cite{Gehrke2019-og, Singh2018-qi, Singh2021-qc, Si-mohammed2020-ru}. This component overlaps in timing and topography with the N2 fronto-central negativity associated with conflict detection and cognitive control~\cite{Folstein2008-ke, Evan_G_Center2025-pk}. Our data reinforce prior work showing that even in movement-rich, immersive VR settings, the PEN can be robustly elicited despite potential motion noise. The absence of significant immersion or presence moderation at FCz suggests that medial-frontal generators of the PEN operate in a more domain-general manner, contradicting other findings of a sensitivity to immersion manipulations~\cite{Gehrke2019-og, Evan_G_Center2025-pk}. One potential reason for this discrepancy are technological differences in the stimulation. In the current study, we explored force-feedback, which has inherent shortcomings in stimulus timing precision as compared to vibrotactile stimulation, where switching on the motor is near instantaneous. This might have resulted in a temporal jitter, smearing the immersion effect.

Nevertheless, the scalp distributions point to additional posterior contributions. The parietal negativity observed from 120~ms onward, together with a brief immersion-related positivity effect around 250~ms, indicates that sensorimotor PE engage broader fronto-parietal networks. The Pz immersion effect is consistent with studies linking parietal potentials to mismatch detection and multisensory integration~\cite{Klimesch2012-ni, Savoie2018-ad}. The topographies also revealed a left-lateralized centro-parietal activity pattern for the immersion effect, likely reflecting the involvement of the right (dominant) hand in grasping and suggesting lateralized somatosensory recruitment with haptic feedback active.

While questionnaire results indicated that self-presence increased with haptic feedback, these inter-individual differences did not systematically moderate ERP effects. Again, we stress that questionnaires capture retrospective impressions, whereas ERPs reflect moment-to-moment responses to specific disruptions. We believe the absence of a strong presence modulation therefore does not necessarily imply a lack of experiential relevance; rather, it underscores the complementary nature of subjective and neural measures of presence.

At FCz, incongruent trials elicited a transient theta burst, consistent with a phase-reset mechanism underlying the ERP negativity~\cite{Cavanagh2014-mm}. No immersion effect was observed at this early stage, again suggesting that theta synchronization reflects generic conflict detection rather than modality-specific processing~\cite{Lin2012-mp}. More intriguingly, we found indications that presence gains through haptic immersion modulated frontal theta during the placement phase, with participants reporting stronger presence showing increased theta power around object release. Theta activity in this later window has been linked to cognitive control and action monitoring~\cite{Gurtubay-Antolin2018-dv, Tollner2017-rm}, and its modulation by presence suggests that a more immersive state heightens the precision demands of future goal-directed actions in VR. This finding echoes previous work in haptic delay paradigms where frontal theta indexed the salience of temporal mismatches~\cite{Alsuradi2022-rw}.

Source-level analyses revealed more nuanced patterns. Here, we did not recover the effect which was present at FCz, in the ACC cluster. While independent component clustering yielded relatively sparse sources in ACC, this may reflect methodological constraints rather than absence of ACC involvement. The clustering approach required strict criteria for component inclusion, potentially reducing statistical power and excluding participants with weaker ACC signals. Future analyses examining sensor-level data with the same participant exclusions may clarify whether this represents a power issue rather than a functional absence of ACC contributions. While ACC EEG sources may be challening to localize, recent works have shown that recovering EEG activity from even deeper cortical layers is possible with advanced source localization techniques~\cite{Fahimi-Hnazaee2020-pg}.

The most striking source-level finding concerned the PCC cluster, where alpha power was selectively suppressed during high-immersion haptic glitches. Posterior alpha desynchronization has been widely linked to enhanced sensory integration and attentional binding~\cite{Klimesch2012-ni}. Our results suggest that the PCC is particularly sensitive to multisensory coherence: increased haptic immersion heightens the impact of sensory mismatches, reflected in stronger alpha suppression. This interpretation is consistent with recent findings showing that body-related multisensory conflicts trigger modality-specific adjustments in visual and somatosensory processing to reestablish coherence~\cite{Rossi_Sebastiano2024-rc, Chancel2022-ih, D-Angelo2025-ap}, as well as with theoretical accounts identifying the PCC as a central hub for self-related multisensory integration~\cite{Tsakiris2017-oe}. Furthermore, first indications of a presence moderation effect manifested as a modulation of the congruency condition following object placement, suggesting that higher presence leads posterior networks to allocate more resources to resolving future incoherence. This timing may reflect the extended processing window required to integrate multisensory feedback and evaluate action outcomes, rather than the immediate PE detection occurring at grasp onset. An alternative explanation is that participants with stronger presence gains exhibited increased alpha rebound after placement. This rebound may reflect a process of closure and re-stabilization following action execution, suggesting that heightened presence not only accentuates the neural costs of mismatches but also modulates recovery dynamics in sensorimotor integration. Complementary findings of sensory attenuation and self-related suppression in VR further support the notion that posterior oscillations play a central role in linking presence, agency, and self-representation~\cite{Giannini2025-dn}.

\subsection{Conclusion}

Our findings provide mixed support for the three hypotheses outlined in the introduction. First, regarding immersion effects on presence, we found partial confirmation: haptic feedback modestly increased self-presence but not physical presence, suggesting that additional sensory cues enhance embodiment while technological limitations constrain environmental realism. Second, PE elicited robust neural signatures at the channel level as predicted, with canonical PEN responses at FCz (170-300ms) and sustained parietal negativities at Pz (120ms onward), confirming that sensorimotor disruptions trigger reliable error monitoring even in complex VR environments. However, at the source level, we could not replicate ACC $\theta$-band involvement, with frontal theta showing domain-general conflict detection rather than context-sensitive processing. The most striking finding was that PCC alpha suppression was selectively enhanced during haptic mismatches, revealing that posterior networks are particularly sensitive to multisensory coherence violations when immersion is high. Lastly, we found no clear, convincing pattern for inter-individual differences in presence questionnaire scores moderate these effects. 

Together, these results suggest a hierarchical model of presence disruption in VR: frontal regions detect prediction violations in a modality-agnostic manner, while posterior networks reflect the contextual costs of incoherence when rich sensory cues amplify coherence demands. This pattern aligns with predictive coding frameworks where frontal monitoring operates independently of sensory immersion, but posterior integration scales with the precision of multisensory predictions. The present study demonstrates that VR environments are particularly well-suited for investigating presence, as multisensory integration dynamics are robustly recoverable using mobile EEG despite movement-related challenges. 

As a final remark, we advocate for the usage of complementary approaches to questionnaires that better capture the subjective experience. These include structured interviews and sentiment analysis which may provide deeper insights into how users experience and interpret multisensory conflicts, bridging the gap between objective neural markers and lived experience in immersive environments.

\bibliographystyle{plainnat}
\bibliography{paperpile}

\begin{thebibliography}{65}
\providecommand{\natexlab}[1]{#1}
\providecommand{\url}[1]{\texttt{#1}}
\expandafter\ifx\csname urlstyle\endcsname\relax
  \providecommand{\doi}[1]{doi: #1}\else
  \providecommand{\doi}{doi: \begingroup \urlstyle{rm}\Url}\fi

\bibitem[Alexander and Brown(2019)]{Alexander2019-uf}
William~H Alexander and Joshua~W Brown.
\newblock The role of the anterior cingulate cortex in prediction error and signaling surprise.
\newblock \emph{Top. Cogn. Sci.}, 11\penalty0 (1):\penalty0 119--135, January 2019.

\bibitem[Alsuradi et~al.(2022)Alsuradi, Park, and Eid]{Alsuradi2022-rw}
Haneen Alsuradi, Wanjoo Park, and Mohamad Eid.
\newblock Midfrontal theta power encodes the value of haptic delay.
\newblock \emph{Sci. Rep.}, 12\penalty0 (1):\penalty0 8869, May 2022.

\bibitem[Benjamini and Hochberg(1995)]{Benjamini1995-cw}
Yoav Benjamini and Yosef Hochberg.
\newblock Controlling the false discovery rate: A practical and powerful approach to multiple testing.
\newblock \emph{J. R. Stat. Soc.}, 57\penalty0 (1):\penalty0 289--300, January 1995.

\bibitem[Bigdely-Shamlo et~al.(2015)Bigdely-Shamlo, Mullen, Kothe, Su, and Robbins]{Bigdely-Shamlo2015-ds}
Nima Bigdely-Shamlo, Tim Mullen, Christian Kothe, Kyung-Min Su, and Kay~A Robbins.
\newblock The {PREP} pipeline: standardized preprocessing for large-scale {EEG} analysis.
\newblock \emph{Front. Neuroinform.}, 9\penalty0 (June):\penalty0 16, June 2015.

\bibitem[Botvinick and Cohen(1998)]{Botvinick1998-iw}
M~Botvinick and J~Cohen.
\newblock Rubber hands 'feel' touch that eyes see.
\newblock \emph{Nature}, 391\penalty0 (6669):\penalty0 756, February 1998.

\bibitem[Bowman et~al.(2012)Bowman, McMahan, and Ragan]{Bowman2012-ga}
Doug~A Bowman, Ryan~P McMahan, and Eric~D Ragan.
\newblock Questioning naturalism in {3D} user interfaces.
\newblock \emph{Commun. ACM}, 55\penalty0 (9):\penalty0 78--88, September 2012.

\bibitem[Cavanagh and Frank(2014)]{Cavanagh2014-mm}
James~F Cavanagh and Michael~J Frank.
\newblock Frontal theta as a mechanism for cognitive control.
\newblock \emph{Trends Cogn. Sci.}, 18\penalty0 (8):\penalty0 414--421, August 2014.

\bibitem[Chancel et~al.(2022)Chancel, Iriye, and Ehrsson]{Chancel2022-ih}
Marie Chancel, Heather Iriye, and H~Henrik Ehrsson.
\newblock Causal inference of body ownership in the posterior parietal cortex.
\newblock \emph{J. Neurosci.}, 42\penalty0 (37):\penalty0 7131--7143, September 2022.

\bibitem[D'Angelo et~al.(2025)D'Angelo, Lanfranco, Chancel, and Ehrsson]{D-Angelo2025-ap}
Mariano D'Angelo, Renzo~C Lanfranco, Marie Chancel, and H~Henrik Ehrsson.
\newblock Parietal alpha frequency shapes own-body perception by modulating the temporal integration of bodily signals.
\newblock \emph{bioRxiv}, February 2025.

\bibitem[Delorme and Makeig(2004)]{Delorme2004-sn}
Arnaud Delorme and Scott Makeig.
\newblock {EEGLAB}: an open source toolbox for analysis of single-trial {EEG} dynamics including independent component analysis.
\newblock \emph{J. Neurosci. Methods}, 134\penalty0 (1):\penalty0 9--21, March 2004.

\bibitem[{Evan G. Center} et~al.(2025){Evan G. Center}, Pouke, Nardi, Gehrke, Gramann, Ojala, and Lavalle]{Evan_G_Center2025-pk}
{Evan G. Center}, Matti Pouke, Alessandro Nardi, L~Gehrke, Klaus Gramann, Timo Ojala, and Steven~M Lavalle.
\newblock Body ownership affects the processing of sensorimotor contingencies in virtual reality.
\newblock September 2025.

\bibitem[Fahimi~Hnazaee et~al.(2020)Fahimi~Hnazaee, Wittevrongel, Khachatryan, Libert, Carrette, Dauwe, Meurs, Boon, Van~Roost, and Van~Hulle]{Fahimi-Hnazaee2020-pg}
Mansoureh Fahimi~Hnazaee, Benjamin Wittevrongel, Elvira Khachatryan, Arno Libert, Evelien Carrette, Ine Dauwe, Alfred Meurs, Paul Boon, Dirk Van~Roost, and Marc~M Van~Hulle.
\newblock Localization of deep brain activity with scalp and subdural {EEG}.
\newblock \emph{Neuroimage}, 223\penalty0 (117344):\penalty0 117344, December 2020.

\bibitem[Feick et~al.(2020)Feick, Kleer, Tang, and Krüger]{Feick2020-fb}
Martin Feick, Niko Kleer, Anthony Tang, and Antonio Krüger.
\newblock The virtual reality questionnaire toolkit.
\newblock In \emph{Adjunct Proceedings of the 33rd Annual ACM Symposium on User Interface Software and Technology}, UIST '20 Adjunct, pages 68--69, New York, NY, USA, October 2020. Association for Computing Machinery.

\bibitem[Folstein and Van~Petten(2008)]{Folstein2008-ke}
Jonathan~R Folstein and Cyma Van~Petten.
\newblock Influence of cognitive control and mismatch on the {N2} component of the {ERP}: a review.
\newblock \emph{Psychophysiology}, 45\penalty0 (1):\penalty0 152--170, January 2008.

\bibitem[Friston(2010)]{Friston2010-hy}
Karl Friston.
\newblock The free-energy principle: a unified brain theory?
\newblock \emph{Nat. Rev. Neurosci.}, 11\penalty0 (2):\penalty0 127--138, February 2010.

\bibitem[Gall and Latoschik(2018)]{Gall2018-pb}
Dominik Gall and Marc~Erich Latoschik.
\newblock The effect of haptic prediction accuracy on presence.
\newblock In \emph{2018 IEEE Conference on Virtual Reality and 3D User Interfaces (VR)}, pages 73--80. IEEE, March 2018.

\bibitem[Gehrke et~al.(2019{\natexlab{a}})Gehrke, Akman, Lopes, Chen, Singh, Chen, Lin, and Gramann]{Gehrke2019-og}
Lukas Gehrke, Sezen Akman, Pedro Lopes, Albert Chen, Avinash~Kumar Singh, Hsiang-Ting Chen, Chin-Teng Lin, and Klaus Gramann.
\newblock Detecting visuo-haptic mismatches in virtual reality using the prediction error negativity of event-related brain potentials.
\newblock In \emph{Proceedings of the 2019 CHI Conference on Human Factors in Computing Systems - CHI '19}, CHI '19, pages 427:1--427:11, New York, New York, USA, 2019{\natexlab{a}}. ACM Press.

\bibitem[Gehrke et~al.(2019{\natexlab{b}})Gehrke, Guerdan, and Gramann]{Gehrke2019-hp}
Lukas Gehrke, Luke Guerdan, and Klaus Gramann.
\newblock Extracting motion-related subspaces from {EEG} in mobile brain/body imaging studies using source power comodulation.
\newblock In \emph{2019 9th International IEEE/EMBS Conference on Neural Engineering (NER)}, pages 344--347, March 2019{\natexlab{b}}.

\bibitem[Gehrke et~al.(2022)Gehrke, Lopes, Klug, Akman, and Gramann]{Gehrke2022-tj}
Lukas Gehrke, Pedro Lopes, Marius Klug, Sezen Akman, and Klaus Gramann.
\newblock Neural sources of prediction errors detect unrealistic {VR} interactions.
\newblock \emph{J. Neural Eng.}, 19\penalty0 (3), May 2022.

\bibitem[Gehrke et~al.(2024)Gehrke, Terfurth, Akman, and Gramann]{Gehrke2024-xq}
Lukas Gehrke, Leonie Terfurth, Sezen Akman, and Klaus Gramann.
\newblock Visuo-haptic prediction errors: a multimodal dataset ({EEG}, motion) in {BIDS} format indexing mismatches in haptic interaction.
\newblock \emph{Front. Neuroergonomics}, 5:\penalty0 1411305, June 2024.

\bibitem[Gehrke et~al.(2025)Gehrke, Koselevs, Klug, and Gramann]{Gehrke2025-ma}
Lukas Gehrke, Aleksandrs Koselevs, Marius Klug, and Klaus Gramann.
\newblock Neuroadaptive haptics: A proof-of-concept comparing reinforcement learning from explicit ratings and neural signals for adaptive {XR} systems.
\newblock \emph{Front. Virtual Real.}, 6:\penalty0 1616442, July 2025.

\bibitem[Giannini et~al.(2025)Giannini, Nierhaus, and Blankenburg]{Giannini2025-dn}
Gianluigi Giannini, Till Nierhaus, and Felix Blankenburg.
\newblock Investigation of sensory attenuation in the somatosensory domain using {EEG} in a novel virtual reality paradigm.
\newblock \emph{Sci. Rep.}, 15\penalty0 (1):\penalty0 2819, January 2025.

\bibitem[Gibo et~al.(2017)Gibo, Mugge, and Abbink]{Gibo2017-gi}
Tricia~L Gibo, Winfred Mugge, and David~A Abbink.
\newblock Trust in haptic assistance: weighting visual and haptic cues based on error history.
\newblock \emph{Exp. Brain Res.}, 235\penalty0 (8):\penalty0 2533--2546, August 2017.

\bibitem[Gonzalez-Franco and Lanier(2017)]{Gonzalez-Franco2017-ff}
Mar Gonzalez-Franco and Jaron Lanier.
\newblock Model of illusions and virtual reality.
\newblock \emph{Front. Psychol.}, 8:\penalty0 1125, June 2017.

\bibitem[Gonzalez-Franco and Peck(2018)]{Gonzalez-Franco2018-jz}
Mar Gonzalez-Franco and Tabitha~C Peck.
\newblock Avatar embodiment. towards a standardized questionnaire, 2018.

\bibitem[Gramann et~al.(2021)Gramann, Hohlefeld, Gehrke, and Klug]{Gramann2021-ug}
Klaus Gramann, Friederike~U Hohlefeld, Lukas Gehrke, and Marius Klug.
\newblock Human cortical dynamics during full-body heading changes.
\newblock \emph{Sci. Rep.}, 11\penalty0 (1):\penalty0 18186, September 2021.

\bibitem[Gramfort et~al.(2013)Gramfort, Luessi, Larson, Engemann, Strohmeier, Brodbeck, Goj, Jas, Brooks, Parkkonen, and Hämäläinen]{Gramfort2013-fa}
Alexandre Gramfort, Martin Luessi, Eric Larson, Denis~A Engemann, Daniel Strohmeier, Christian Brodbeck, Roman Goj, Mainak Jas, Teon Brooks, Lauri Parkkonen, and Matti Hämäläinen.
\newblock {MEG} and {EEG} data analysis with {MNE}-python.
\newblock \emph{Front. Neurosci.}, 7:\penalty0 267, December 2013.

\bibitem[Gurtubay-Antolin et~al.(2018)Gurtubay-Antolin, León-Cabrera, and Rodríguez-Fornells]{Gurtubay-Antolin2018-dv}
Ane Gurtubay-Antolin, Patricia León-Cabrera, and Antoni Rodríguez-Fornells.
\newblock Neural evidence of hierarchical cognitive control during haptic processing: An {fMRI} study.
\newblock \emph{eNeuro}, 5\penalty0 (6):\penalty0 ENEURO.0295--18.2018, November 2018.

\bibitem[Holroyd and Coles(2002)]{Holroyd2002-in}
Clay~B Holroyd and Michael G~H Coles.
\newblock The neural basis of human error processing: reinforcement learning, dopamine, and the error-related negativity.
\newblock \emph{Psychol. Rev.}, 109\penalty0 (4):\penalty0 679--709, October 2002.

\bibitem[Jolly(2018)]{Jolly2018-it}
Eshin Jolly.
\newblock {Pymer4}: Connecting {R} and python for linear mixed modeling.
\newblock \emph{J. Open Source Softw.}, 3\penalty0 (31):\penalty0 862, November 2018.

\bibitem[Kilteni et~al.(2012)Kilteni, Groten, and Slater]{Kilteni2012-gm}
Konstantina Kilteni, Raphaela Groten, and Mel Slater.
\newblock The sense of embodiment in virtual reality, 2012.

\bibitem[Klimesch(2012)]{Klimesch2012-ni}
Wolfgang Klimesch.
\newblock α-band oscillations, attention, and controlled access to stored information.
\newblock \emph{Trends Cogn. Sci.}, 16\penalty0 (12):\penalty0 606--617, December 2012.

\bibitem[Klug et~al.(2022)Klug, Jeung, Wunderlich, Gehrke, Protzak, Djebbara, Argubi-Wollesen, Wollesen, and Gramann]{Klug2022-lc}
M~Klug, S~Jeung, A~Wunderlich, L~Gehrke, J~Protzak, Z~Djebbara, A~Argubi-Wollesen, B~Wollesen, and K~Gramann.
\newblock The {BeMoBIL} pipeline for automated analyses of multimodal mobile brain and body imaging data.
\newblock \emph{bioRxiv}, page 2022.09.29.510051, October 2022.

\bibitem[Kothe et~al.(2025)Kothe, Shirazi, Stenner, Medine, Boulay, Grivich, Artoni, Mullen, Delorme, and Makeig]{Kothe2025-oh}
Christian Kothe, Seyed~Yahya Shirazi, Tristan Stenner, David Medine, Chadwick Boulay, Matthew~I Grivich, Fiorenzo Artoni, Tim Mullen, Arnaud Delorme, and Scott Makeig.
\newblock The lab streaming layer for synchronized multimodal recording.
\newblock \emph{Imaging Neurosci. (Camb.)}, 3:\penalty0 IMAG. a. 136, September 2025.

\bibitem[Krol et~al.(2020)Krol, Haselager, and Zander]{Krol2020-lj}
Laurens~R Krol, Pim Haselager, and Thorsten~O Zander.
\newblock Cognitive and affective probing: a tutorial and review of active learning for neuroadaptive technology.
\newblock \emph{J. Neural Eng.}, 17\penalty0 (1):\penalty0 012001, January 2020.

\bibitem[Körding et~al.(2007)Körding, Beierholm, Ma, Quartz, Tenenbaum, and Shams]{Kording2007-xf}
Konrad~P Körding, Ulrik Beierholm, Wei~Ji Ma, Steven Quartz, Joshua~B Tenenbaum, and Ladan Shams.
\newblock Causal inference in multisensory perception.
\newblock \emph{PLoS One}, 2\penalty0 (9):\penalty0 e943, September 2007.

\bibitem[Limanowski(2022)]{Limanowski2022-ds}
Jakub Limanowski.
\newblock Precision control for a flexible body representation.
\newblock \emph{Neurosci. Biobehav. Rev.}, 134:\penalty0 104401, March 2022.

\bibitem[Lin et~al.(2012)Lin, Shaw, Young, Lin, and Jung]{Lin2012-mp}
Chun-Ling Lin, Fu-Zen Shaw, Kuu-Young Young, Chin-Teng Lin, and Tzyy-Ping Jung.
\newblock {EEG} correlates of haptic feedback in a visuomotor tracking task.
\newblock \emph{Neuroimage}, 60\penalty0 (4):\penalty0 2258--2273, May 2012.

\bibitem[Makeig et~al.(1995)Makeig, Bell, Jung, and Sejnowski]{Makeig1995-cf}
S~Makeig, A~J Bell, T~Jung, and T~Sejnowski.
\newblock Independent component analysis of electroencephalographic data.
\newblock \emph{Neural Inf Process Syst}, 8:\penalty0 145--151, November 1995.

\bibitem[Makin et~al.(2007)Makin, Holmes, and Zohary]{Makin2007-ng}
Tamar~R Makin, Nicholas~P Holmes, and Ehud Zohary.
\newblock Is that near my hand? multisensory representation of peripersonal space in human intraparietal sulcus.
\newblock \emph{J. Neurosci.}, 27\penalty0 (4):\penalty0 731--740, January 2007.

\bibitem[Makransky et~al.(2017)Makransky, Lilleholt, and Aaby]{Makransky2017-hr}
Guido Makransky, Lau Lilleholt, and Anders Aaby.
\newblock Development and validation of the multimodal presence scale for virtual reality environments: A confirmatory factor analysis and item response theory approach.
\newblock \emph{Comput. Human Behav.}, 72:\penalty0 276--285, July 2017.

\bibitem[Michel and Brunet(2019)]{Michel2019-bb}
Christoph~M Michel and Denis Brunet.
\newblock {EEG} source imaging: A practical review of the analysis steps.
\newblock \emph{Front. Neurol.}, 10, April 2019.

\bibitem[Misselhorn et~al.(2019)Misselhorn, Friese, and Engel]{Misselhorn2019-bs}
Jonas Misselhorn, Uwe Friese, and Andreas~K Engel.
\newblock Frontal and parietal alpha oscillations reflect attentional modulation of cross-modal matching.
\newblock \emph{Sci. Rep.}, 9\penalty0 (1):\penalty0 5030, March 2019.

\bibitem[Oostenveld and Praamstra(2001)]{Oostenveld2001-yd}
R~Oostenveld and P~Praamstra.
\newblock The five percent electrode system for high-resolution {EEG} and {ERP} measurements.
\newblock \emph{Clin. Neurophysiol.}, 112\penalty0 (4):\penalty0 713--719, April 2001.

\bibitem[Palmer et~al.(2011)Palmer, Kreutz-Delgado, and Makeig]{Palmer2011-zs}
Jason Palmer, Ken Kreutz-Delgado, and Scott Makeig.
\newblock {AMICA}: An adaptive mixture of independent component analyzers with shared components.
\newblock \emph{San Diego, CA: Technical report, Swartz Center for Computational Neuroscience}, pages 1--15, 2011.

\bibitem[Pinheiro and Bates(2006)]{Pinheiro2006-bk}
José Pinheiro and Douglas Bates.
\newblock \emph{Mixed-Effects Models in {S} and {S}-{PLUS}}.
\newblock Springer Science \& Business Media, May 2006.

\bibitem[Pion-Tonachini et~al.(2019)Pion-Tonachini, Kreutz-Delgado, and Makeig]{Pion-Tonachini2019-fy}
Luca Pion-Tonachini, Ken Kreutz-Delgado, and Scott Makeig.
\newblock {ICLabel}: An automated electroencephalographic independent component classifier, dataset, and website.
\newblock \emph{Neuroimage}, 198:\penalty0 181--197, September 2019.

\bibitem[Rossi~Sebastiano et~al.(2024)Rossi~Sebastiano, Poles, Gualtiero, Romeo, Galigani, Bruno, Fossataro, and Garbarini]{Rossi_Sebastiano2024-rc}
Alice Rossi~Sebastiano, Karol Poles, Stefano Gualtiero, Marcella Romeo, Mattia Galigani, Valentina Bruno, Carlotta Fossataro, and Francesca Garbarini.
\newblock Balancing the senses: Electrophysiological responses reveal the interplay between somatosensory and visual processing during body-related multisensory conflict.
\newblock \emph{J. Neurosci.}, 44\penalty0 (19):\penalty0 e1397232024, May 2024.

\bibitem[Savalle et~al.(2024)Savalle, Pillette, Won, Argelaguet, Lécuyer, and J-M~Macé]{Savalle2024-co}
Emile Savalle, Léa Pillette, Kyungho Won, Ferran Argelaguet, Anatole Lécuyer, and Marc J-M~Macé.
\newblock Towards electrophysiological measurement of presence in virtual reality through auditory oddball stimuli.
\newblock \emph{J. Neural Eng.}, 21\penalty0 (4):\penalty0 046015, July 2024.

\bibitem[Savoie et~al.(2018)Savoie, Thénault, Whittingstall, and Bernier]{Savoie2018-ad}
F-A Savoie, F~Thénault, K~Whittingstall, and P-M Bernier.
\newblock Visuomotor prediction errors modulate {EEG} activity over parietal cortex.
\newblock \emph{Sci. Rep.}, 8\penalty0 (1):\penalty0 12513, August 2018.

\bibitem[Schubert(2003)]{Schubert2003-sq}
Thomas~W Schubert.
\newblock The sense of presence in virtual environments:.
\newblock \emph{Zeitschrift für Medienpsychologie}, 15\penalty0 (2):\penalty0 69--71, April 2003.

\bibitem[Schwind et~al.(2019)Schwind, Knierim, Haas, and Henze]{Schwind2019-ar}
Valentin Schwind, Pascal Knierim, Nico Haas, and Niels Henze.
\newblock Using presence questionnaires in virtual reality.
\newblock In \emph{Proceedings of the 2019 CHI Conference on Human Factors in Computing Systems}, number Paper 360 in CHI '19, pages 1--12, New York, NY, USA, May 2019. Association for Computing Machinery.

\bibitem[Seth(2014)]{Seth2014-fv}
Anil~K Seth.
\newblock A predictive processing theory of sensorimotor contingencies: Explaining the puzzle of perceptual presence and its absence in synesthesia.
\newblock \emph{Cogn. Neurosci.}, 5\penalty0 (2):\penalty0 97--118, January 2014.

\bibitem[Seth and Hohwy(2021)]{Seth2021-io}
Anil~K Seth and Jakob Hohwy.
\newblock Predictive processing as an empirical theory for consciousness science.
\newblock \emph{Cogn. Neurosci.}, 12\penalty0 (2):\penalty0 89--90, January 2021.

\bibitem[Si-mohammed et~al.(2020)Si-mohammed, Lopes-dias, Duarte, Jeunet, and Scherer]{Si-mohammed2020-ru}
Hakim Si-mohammed, Catarina Lopes-dias, Maria Duarte, Camille Jeunet, and Reinhold Scherer.
\newblock Detecting system errors in virtual reality using {EEG} through error-related potentials.
\newblock \penalty0 (1):\penalty0 653--661, 2020.

\bibitem[Singh et~al.(2021)Singh, Gramann, Chen, and Lin]{Singh2021-qc}
Avinash~K Singh, Klaus Gramann, Hsiang-Ting Chen, and Chin-Teng Lin.
\newblock The impact of hand movement velocity on cognitive conflict processing in a {3D} object selection task in virtual reality.
\newblock \emph{Neuroimage}, 226\penalty0 (April 2020):\penalty0 117578, February 2021.

\bibitem[Singh et~al.(2018)Singh, Chen, Cheng, King, Ko, Gramann, and Lin]{Singh2018-qi}
Avinash~Kumar Singh, Hsiang-Ting Chen, Yu-Feng Cheng, Jung-Tai King, Li-Wei Ko, Klaus Gramann, and Chin-Teng Lin.
\newblock Visual appearance modulates prediction error in virtual reality.
\newblock \emph{IEEE Access}, 6:\penalty0 24617--24624, 2018.

\bibitem[Slater et~al.(2003)Slater, Brogni, and Steed]{Slater2003-gg}
M~Slater, A~Brogni, and A~Steed.
\newblock Physiological responses to breaks in presence: A pilot study.
\newblock 2003.

\bibitem[Slater(2009)]{Slater2009-au}
Mel Slater.
\newblock Place illusion and plausibility can lead to realistic behaviour in immersive virtual environments.
\newblock \emph{Philos. Trans. R. Soc. Lond. B Biol. Sci.}, 364\penalty0 (1535):\penalty0 3549--3557, December 2009.

\bibitem[Slater et~al.(1994)Slater, Usoh, and Steed]{Slater1994-sh}
Mel Slater, Martin Usoh, and Anthony Steed.
\newblock Depth of presence in virtual environments.
\newblock \emph{Presence (Camb.)}, 3\penalty0 (2):\penalty0 130--144, January 1994.

\bibitem[Tsakiris(2017)]{Tsakiris2017-oe}
Manos Tsakiris.
\newblock The multisensory basis of the self: From body to identity to others.
\newblock \emph{Q. J. Exp. Psychol.}, 70\penalty0 (4):\penalty0 597--609, April 2017.

\bibitem[Töllner et~al.(2017)Töllner, Wang, Makeig, Müller, Jung, and Gramann]{Tollner2017-rm}
Thomas Töllner, Yijun Wang, Scott Makeig, Hermann~J Müller, Tzyy-Ping Jung, and Klaus Gramann.
\newblock Two independent frontal midline theta oscillations during conflict detection and adaptation in a simon-type manual reaching task.
\newblock \emph{J. Neurosci.}, 37\penalty0 (9):\penalty0 2504--2515, March 2017.

\bibitem[Van~Noordt et~al.(2016)Van~Noordt, Campopiano, and Segalowitz]{Van_Noordt2016-qz}
Stefon J~R Van~Noordt, Allan Campopiano, and Sidney~J Segalowitz.
\newblock A functional classification of medial frontal negativity {ERPs}: Theta oscillations and single subject effects.
\newblock \emph{Psychophysiology}, 53\penalty0 (9):\penalty0 1317--1334, September 2016.

\bibitem[Weech et~al.(2019)Weech, Kenny, and Barnett-Cowan]{Weech2019-pm}
Séamas Weech, Sophie Kenny, and Michael Barnett-Cowan.
\newblock Presence and cybersickness in virtual reality are negatively related: A review.
\newblock \emph{Front. Psychol.}, 10:\penalty0 158, February 2019.

\bibitem[Witmer and Singer(1998)]{Witmer1998-ew}
Bob~G Witmer and Michael~J Singer.
\newblock Measuring presence in virtual environments: A presence questionnaire.
\newblock \emph{Presence}, 7\penalty0 (3):\penalty0 225--240, June 1998.

\end{thebibliography}

\appendix


\section{Outlier Removal}

Grab duration histograms for each participant, vertical lines mark established outlier bounds.

\begin{figure*}[h]
  \includegraphics[width=.8\textwidth]{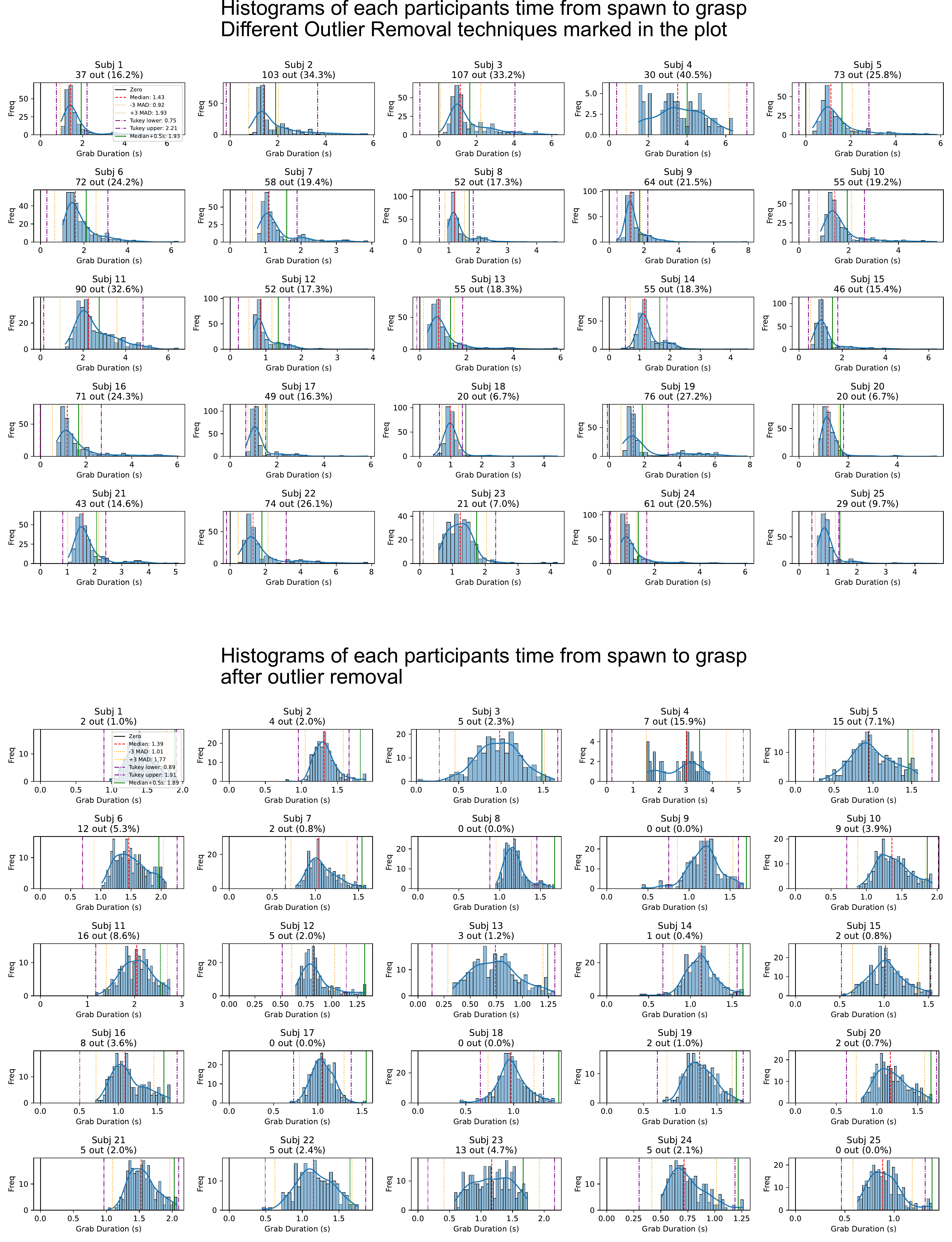}
  \caption{Outlier removal procedure for grab duration. (\textbf{Top}) Histograms of trial-wise grab durations for each participant before outlier removal. Vertical lines indicate thresholds based on multiple detection methods (median $\pm$3 MAD, Tukey fences, median $+0.5$~s). The number and percentage of excluded trials per participant are shown. (\textbf{Bottom}) Histograms of grab durations after outlier removal, illustrating the retained distributions.}
  \label{fig:outlier1}
\end{figure*}

\end{document}